\documentclass[aps,amsfonts,amsmath,prd,tightenlines,twocolumn,showpacs,nofootinbib]{revtex4}
\usepackage{epsf}

\def\be{\begin{equation}}
\def\ee{\end{equation}}
\def\beq{\begin{equation}}
\def\eeq{\end{equation}}
\def\bea{\begin{eqnarray}}
\def\eea{\end{eqnarray}}
\def\bml{\begin{subequations}}
\def\blea{\bml\begin{eqnarray}}
\def\elea{\end{eqnarray}\end{subequations}}

\def\bS{\mathbf{S}}
\def\bD{\mathbf{D}}
\def\bX{\mathbf{X}}
\def\bx{\mathbf{x}}
\def\bt{\mathbf{t}}
\def\bT{\mathbf{T}}
\def\bM{\mathbf{M}}
\def\bN{\mathbf{N}}
\def\bI{\mathbf{I}}
\def\bR{\mathbf{R}}
\def\bp{\mathbf{p}}

\begin{document}

\title{Geodesic measures of the landscape}

\author{Vitaly Vanchurin}
\email{vitaly@cosmos.phy.tufts.edu}

\affiliation{Arnold-Sommerfeld-Center for Theoretical Physics, Department f\"ur Physik, 
Ludwig-Maximilians-Universit\"at M\"unchen, Theresienstr. 37, D-80333, Munich, Germany}

\begin{abstract}

We study the landscape models of eternal inflation with an arbitrary number of different vacua states, both recyclable and terminal. We calculate the abundances of bubbles following different geodesics. We show that the results obtained from generic time-like geodesics have dependence on initial conditions. In contrast, the predictions extracted from ``eternal'' geodesics, which never enter terminal vacua, do not suffer from this problem.  We derive measure equations for ensembles of geodesics and discuss possible interpretations of initial conditions in eternal inflation.

\end{abstract}

\maketitle 

\section{Introduction}

Models of inflation generically lead to space-times with an infinite number of eternally inflating domains \cite{Vilenkin, Linde}. The eternal self-reproduction of inflating regions, gives birth to the multiverse containing an infinite number of different universes. One of the greatest challenges in modern cosmology is to calculate the probability distribution of observables within such multiverse. Early attempts showed a very strong dependence of the result on the cutoff procedure that one imposes \cite{LLM}. 

A gauge-invariant regularization method for models with continuous variation of observable parameters was developed using a spherical cutoff procedure \cite{AV, VVW}. However, recent advances in string theory led to the landscape picture of universe containing a very large number of distinct vacua  ($N \sim 10^{500}$). To address the problem a pocket-based measure was developed \cite{GPVW}. A key ingredient of the proposed method is based on calculations of relative abundances of the bubble universes of different types, which is the main subject of this article.

In Ref. \cite{GPVW} the comoving horizon (CH) cutoff procedure was introduced for measuring the abundances of bubbles. As an alternative procedure of Ref. \cite{ELM} was proved to be analogous to CH method \cite{GPVW}, we will not distinguish between these two measures. 

For models of eternal inflation, in which all vacua have positive energy density,  the worldline (WL) cutoff method was recently introduced \cite{VV}. The idea is to follow the worldline of a single observer and to define the abundances of  bubbles as the frequency at which bubbles of a certain type are visited. Surprisingly, CH and WL methods give identical results for models with full recycling. Nevertheless, WL procedure seems to be more intuitive, as it is derived from the point of view of a single observer.

A natural extension of the WL method to models with terminal vacua, which we will refer to as the Holographic (HG) measure, was discussed in  Ref.  \cite{Bousso}\footnote{The methods of Refs. \cite{VV} and \cite{Bousso} were first described in \cite{Jaume}.}. It was shown that the HG prescription has a strong dependence on initial conditions, which was argued to be given by quantum cosmology. 

In this article we would like to take a somewhat different approach in order to extend WL procedure to models with terminal vacua. We will preserve the most important property of WL and will study the space-time structure of the universe following different geodesics.

In the second section we define two different classes of geodesic measures. The first class of measures is obtained from a single geodesic, which is analyzed in the third section.  The second class of measures is defined from an ensemble of geodesics, which is discussed in the forth section. All of the proposed measures are compared with  CH method in the fifth section for a toy model. In the last section, we conclude with remarks to some recent publications.

\section{Proposals}

\begin{figure}
\begin{center}
\leavevmode\epsfxsize=3.0in\epsfbox{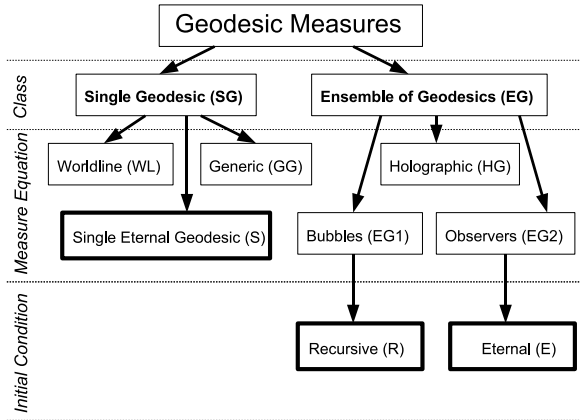}
\end{center}
\caption{Classification of Geodesic Measures.}
\label{fig:measures}
\end{figure}

There are two different ways to assign probabilities in the landscape models, which form two distinct classes of geodesic measures.  Classification diagram of geodesic measures is shown on Fig.\ \ref{fig:measures}
. In this section we will define both classes and propose three measures of the landscape.

\subsection{Single geodesic}

The first approach is to choose a single geodesic and calculate the distribution of bubbles of different types $\bp_{sg}$ that are crossed by the geodesic. We will refer to this class of methods as the Single Geodesic (SG) measures. For example, WL method of Ref. \cite{VV} is of SG class. Our aim is to define a measure of SG class by extending WL method to models with terminal vacua.

Let us choose an arbitrary space-time point $O$, which is in some recyclable vacuum. Although, most of future time-like geodesics passing though $O$ end up in terminal vacua, there are some geodesics that forever remain in recyclable vacua. We call such geodesics eternal, as they are related to the concept of eternal points discussed in Ref. \cite{Serge}. 

\begin{figure}
\begin{center}
\leavevmode\epsfxsize=3.0in\epsfbox{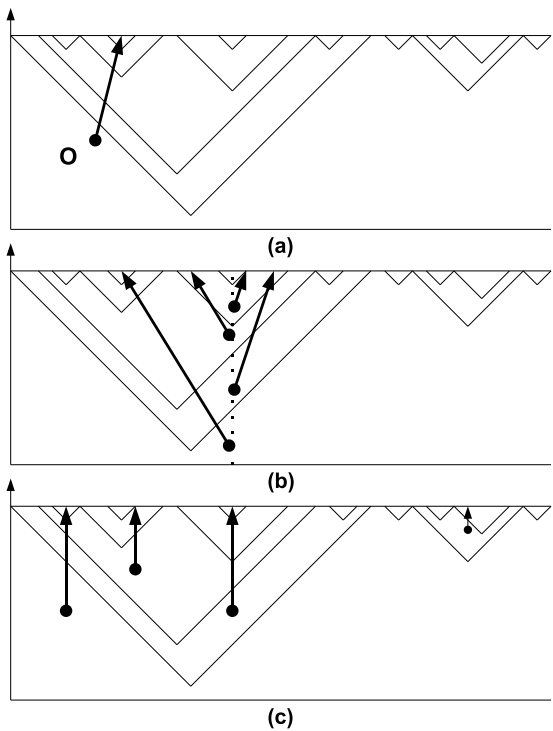}
\end{center}
\caption{Conformal diagrams of the proposed measures: (a) Single eternal geodesic measure, (b) Eternal measure and (c) Recursive measure.}
\label{fig:diagram}
\end{figure}

An eternal geodesic passing through $O$ generally intersects an infinite number of recyclable bubbles of all possible types  (Fig.\ \ref{fig:diagram}
a). Therefore, by following a single eternal geodesic one can define the abundances of bubbles as the frequency of visits to different vacua. This is the first and only measure of the SG class, which is independent of initial conditions. We will refer to this measure as the S measure.  (See Appendix for discussion of the existence and uniqueness of S measure).

\subsection{Ensemble of geodesics}

Geodesic measures of the second type $\bp_{eg}$ are calculated from an ensemble of geodesics.  We will denote the methods of this class as the Ensemble of Geodesics (EG) measures. For example, the HG measure is of EG class.

The measures of EG class are calculated with two necessary ingredients. One has to choose an ensemble of  observers, and calculate the relative abundances of bubbles from the point of view of these observer. However, this is not sufficient to define a measure, unless we specify how different observers or different bubbles should be weighted.

There are two distinct prescriptions that one can adopt: either all of the observers are equivalent ($\bp_{eg2}$), or all of the bubbles are equivalent ($\bp_{eg1}$). These two assumptions give different results, and thereby one should clearly state which prescription is used for calculations.

The outcome of EG methods could have a strong dependence on initial conditions. One can either specify the initial distribution of observers, or postulate a rule, which determines initial conditions. In what follows, we will study one initial distribution, generated by an eternal geodesic (dotted line on Fig.\ \ref{fig:diagram} b), and one rule, which we will refer to as the recursive principle (Fig.\ \ref{fig:diagram} c): {\it The measure is determined by an ensemble of geodesics with initial conditions given by the measure itself.} In the forth section, the eternal and recursive initial conditions will be explained in greater details. 

\section{Time-like geodesics}

Bubbles intersected by different time-like geodesics starting from a given origin $O$ are generated by the same Markov process.  Therefore, the expected abundance of bubbles $\bp_{gg}$ along different geodesics is also the same (the subscript $gg$ stands for the generic geodesic measure). It was previously calculated for models with only recyclable vacua \cite{VV} and for  models with both, recyclable and terminal vacua \cite{Bousso}.  

\subsection{Generic geodesic}

We define $T_{ij}$ as a probability to make a transition to vacuum $i$ starting from vacuum $j$:
\be
T_{ij} = \begin{cases} \frac{\kappa_{ij}}{\sum_{k=1}^N \kappa_{kj}} & \text{if}\, j \, \text{is recyclable}\\
0 &  \text{if}\, j \, \text{is terminal}
\end{cases}
\ee
where $\kappa_{ij}$ is the tunneling rate from vacuum $j$ to vacuum $i$. 

For models with terminal vacua, we can also define $S_{ij}$, as an expected number of visits to vacuum $i$ starting from vacuum $j$.  From this definition it is clear that the elements of $\bS$ must satisfy the following equation:
\be
S_{ij} = \delta_{ij} + \sum_{k=1}^N S_{ik} T_{kj}.
\ee
The solution of the matrix equation is given by:
\be
\bS=(\bI-\bT)^{-1}.
\ee
The determinant of $\bI-\bT$ is non-zero and the inverse matrix exists, as it was shown in Ref.\cite{Bousso}. We conclude, that the abundances of bubbles are given by:
\be
\bp_{gg} \propto \bS \bp_0  =   (\bI-\bT)^{-1} \bp_0,
\ee
where $\bp_0$ is the state in the origin. If the origin is in $j$'th vacuum, then the $j$'th element of $\bp_0$ is one, and all other elements are exactly zero. We will generalize $\bp_0$ to describe the distribution of observers in the following section.

\subsection{Eternal geodesic}

As one follows a generic eternal geodesic,  the transitions between different vacua are described by Markov process with transition matrix $\bR$ given by:
\be
\bR = \frac{\tilde{\kappa}_{ij}}{\sum_{k=1}^r \tilde{\kappa}_{kj}},
\ee
where  $\tilde{\kappa}$ is a truncated $\kappa$ matrix consisting of only recyclable vacua. A stationary distribution  $\bp_{s}$ for this process is described by a unique solution independent of initial conditions. The solution is given by an eigenvector with eigenvalue one:
\be
\bR \bp_{s} =  \bp_{s}.
\ee

The S measure given by vector $\bp_{s}$ is the first of three measures, which we are proposing in this article. It is independent of initial conditions and almost surely independent of eternal geodesic (See Appendix).

Although it is possible that $\bp_{s}$ is the measure of the universe, it has a feature that might be undesirable.  The vector  $\bp_{s}$ discriminates terminal vacua by assigning them a zero probability. This agrees with observations (we live in a recyclable vacuum), but it was not derived from the first principles, and thus we will continue further our search for the correct measure.

\section{Ensemble of geodesics}

By looking at a single eternal geodesic we are able to get an unambiguous measure of the landscape independent of initial conditions. An alternative approach is to take an ensemble of space-time points given by some distribution $\bp_0$, and follow the generic geodesics starting at these points.

\subsection{Measure equations}

Expected number of bubbles intersected by each geodesic can vary depending on the state of the initial bubble. This suggests two distinct prescriptions for measuring bubble abundances for a given ensemble. The first approach is to assume that all bubbles crossed by geodesics are equally important. In other words, one has to walk along these geodesics one by one and record all of the visited bubbles. As the number of bubbles on each generic geodesic is finite, the total number of counted bubbles can be unambiguously taken to infinity.

An alternative prescription is to calculate the abundances of bubbles measured by the observer on each geodesic separately and then find the abundances averaged over all observers. The distribution  $\bp_{eg1}$ obtained by the former approach treats all of the bubbles with equal weight, while the latter distribution  $\bp_{eg2}$ assumes that all observers are equally important.

It is convenient to define a normalization operator $\bN[\bX]$ acting on a matrix $\bX$, which normalizes each column of $\bX$ to one, if the sum of all elements in the column is non-zero ({\it e.g.} $\bT=\bN[\kappa]$ and $\bR=\bN[{\tilde \kappa}]$). The operator $\bN[\bx]$ can also be used to normalize the components of a column vector $\bx$. For example, the generic geodesic measure is given by
\be
\bp_{gg}=\bN[(\bI-\bT)^{-1} \bp_0].
\ee

Both of the measures ($\bp_{eg1}$ and $\bp_{eg2}$) could be derived from the expression for $\bp_{gg}$:
\be
\label{eg1}
\bp_{eg1} = \bN[(\bI-\bT)^{-1} \bp_0]
\ee
and
\be
\label{eg2}
\bp_{eg2} = \bN[(\bI-\bT)^{-1}] \bp_0.
\ee
Note, that in Eq.~\ref{eg1} the normalization operator acts on a column vector, while in Eq.~\ref{eg2} the operator acts on a square matrix. In former case the operator can be replaced by some multiplicative constant, while in the latter case it could be described by a diagonal matrix. 

One should compare Eqs.~\ref{eg1} and \ref{eg2} with HG measure of Ref. \cite{Bousso}. In our notations the measure is given by:
\be
\bp_{hg}  =  \bN[(\bI-\bT)^{-1} \bT \bp_0],
\ee
and it is obvious that all three expressions could lead to very different results. 

It appears that $\bp_{hg} $ gives us the distribution of bubbles visited by observers assuming that all bubbles  are equally important, but not counting the original bubble. The prescription  $\bp_{hg}$ (as well as $\bp_{eg1}$) discriminates the observers visiting few bubbles over the observers that visit many bubbles. Also, the original bubbles are completely ignored by $\bp_{hg}$ and do not contribute to the final probability distribution.

If different observers can cross the same bubble, the bubble counting measures $\bp_{eg1}$ and $\bp_{hg}$ are somewhat problematic.\footnote{The problem of over-counting was fixed for CH measure discussed in Ref. \cite{ELM} by disregarding all but one world-line that ends up in the same bubble.}  For such ensembles a much more reasonable measure is based on the assumption that all observers are equally important, which is given by $\bp_{eg2}$. On the other hand, if the observers do not pass through the same bubble, then one should probably use $\bp_{eg1}$ in order to weight all bubbles equally.

\subsection{Initial conditions}

As we have proposed, one has to adopt some principle in order to define initial conditions. Based on the principle and on some measure equation unambiguous prescription can be formulated. For example, in Ref. \cite{Bousso} the initial distribution $\bp_0$ was assumed to be given by quantum cosmology. In what follows we will discuss two alternative prescriptions.

Every bubble in eternal inflation is created in the vicinity of some eternal geodesic. Since all of the eternal geodesics are statistically the same (See Appendix), we can choose a single eternal geodesic and treat it as initial conditions (Fig.\ \ref{fig:diagram} b). The first proposal is to picks an arbitrary eternal geodesic, and then to choose generic geodesics which start on the eternal geodesic that we peaked, with only one generic geodesic originated at each bubble intersected by the eternal geodesic. This creates an ensemble of geodesics, which we will refer to as the eternal ensemble. As the distribution of bubbles on a single eternal geodesic $\bp_{s}$ was already derived, the abundances of bubbles measured by the eternal ensemble are given by 
\be
\bp_e = \bN[(\bI-\bT)^{-1}] \bp_{s},
\ee
where the subscript stands for the eternal measure. The measure of Eq.~\ref{eg2} was used in order to treat different observers equally. 

Another way to look at the problem is to adopt the recursive principle. The idea is to assume that the initial conditions in eternally inflating space-time are given by distribution of bubbles in the universe. In other words, the final distribution of bubbles, which we are trying to find, is nothing but the initial distribution . Recursive measure based on the recursive principle could be found from the following equation:
\be
\bp_r = \bN[(\bI-\bT)^{-1} \bp_r].
\ee
where the measure of Eq.~\ref{eg1} was used. The equation can be rewritten as:
\be
\bT \bp_r = \lambda \bp_r,
\ee
where $\bp_r$ is an eigenvector of $\bT$ with eigenvalue $\lambda$. Physically interesting values of $\lambda$ must be real numbers between 0 and  1. It can be shown that for a stochastic matrix $\bT$ with irreducible recyclable vacua and some terminal vacua, there always exists the dominant eigenvalue between 0 and 1 which is non-degenerate \cite{Matrices}. 

Note, that the two measures discussed in this section, are just examples of the measures of EG class. Variations of both eternal $\bp_e$ and recursive $\bp_r$ measures could be formulated by applying different measure equations. 

\section{Examples}

Consider a simple model with three vacua: one terminal and two identical recyclable, for which the decay matrix is given by:
\be
\kappa = \begin{pmatrix}
0 & 1/2 & 0 \\
1/2 & 0 &0 \\
1/2 & 1/2 & 0 
\end{pmatrix}
\ee
\begin{figure}
\begin{center}
\leavevmode\epsfxsize=3.0in\epsfbox{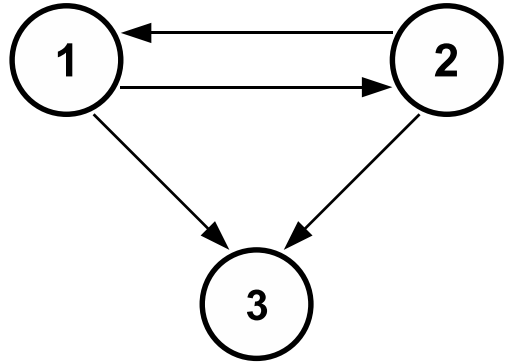}
\end{center}
\caption{A schematic diagram of the model with three vacua: two recyclable (1 and 2) and one terminal (3).}
\label{fig:example}
\end{figure}
Allowed transitions are shown by a schematic diagram (Fig.\ \ref{fig:example}). In what follows, we will calculate the abundances of bubbles measured by four different prescriptions:

- CH measure ($\bp_{ch}$),

- S measure ($\bp_s$),

- Eternal measure ($\bp_e$),

- Recursive measure ($\bp_r$).

In order to apply the CH procedure, one has to calculate the matrix of Ref. \cite{GPVW}:
\be
\bM = \kappa - \bD
\ee
where $\bD$ is a diagonal matrix with $D_{ii}= \sum_{k=1}^N \kappa_{ki}$. In our example
\be 	
\bM = \begin{pmatrix}
-1 & 1/2 & 0 \\
1/2 & -1 &0 \\
1/2 & 1/2 & 0 
\end{pmatrix}.
\ee
The smallest in magnitude negative eigenvalue of $\bM$ is -1/2 and corresponding eigenvector is (1, 1, -2). If $H_1=H_2$, then using the expression of Ref. \cite{GPVW}, the abundances of bubbles are given by:
\be
\bp_c=(1/4, 1/4, 1/2)^\bt.
\ee

For three other prescriptions we need transition matrices
\be
\bT  = \begin{pmatrix}
0 & 1/2 & 0 \\
1/2 & 0 &0 \\
1/2 & 1/2 & 0 
\end{pmatrix}
\ee
and 
\be
\bR =  \begin{pmatrix}
0 & 1/2 \\
1/2 & 0  
\end{pmatrix}.
\ee
Therefore, the S measure is given by the following vector:
\be
\bp_{s} = (1/2,1/2, 0)^\bt.
\ee

The inverse matrix is:
\be
(\bI-\bT)^{-1} = \begin{pmatrix}
4/3 & 2/3 & 0 \\
2/3 & 4/3 & 0 \\
1 & 1 & 1 
\end{pmatrix}
\ee
and the distribution vector of the eternal measure is given by:
\be
\bp_e = (1/3, 1/3, 1/3)^\bt.
\ee

The eigenvector of $\bT$ determines the recursive measure
\be
\bp_r = (1/4, 1/4, 1/2)^\bt,
\ee
which agrees with the result of CH procedure. 

In general this must not be the case, since the recursive measure has no explicit dependence on $H$, unlike the CH measure, which is proportional to $H^q$.  However, when $q$ is very small, one can neglect the $H$ term, and then the CH measure is given by $\kappa$ multiplied by an eigenvector of $\bM$. Therefore, the vector $\bp_{ch}$ is nothing but an eigenvector of
\be
\tilde{\bM}=\kappa \bM \kappa^{-1}= \kappa (\kappa - \bD) \kappa^{-1}= \kappa - \kappa \bD \kappa^{-1},
\ee
while $\bp_{r}$ is an eigenvector of $\bT=\bN[\kappa]$. Thus, it is clear that CH and recursive measures are not always equivalent.

\section{Conclusion}

In this article, we have discussed two distinct classes of geodesic measure and proposed three measures of the universe. The eternal and recursive measures ($\bp_e$ and $\bp_r$) involve two different ensembles of observers, while the measure $\bp_{s}$ is defined by a single eternal observer. 

The probability distribution of bubbles is obtained by their abundances measured along different geodesics ($\bp_{s}, \bp_e$ and $\bp_r$). However, the vector $\bp$ is only one of the ingredients in the complete probability measure. In order to calculate the probability of finding ourselves in a certain vacuum, the full probability measure has to be used (See Ref.\cite{GPVW} for details.)

A number of recent papers  \cite{Page, Freivogel, Linde2, Vilenkin2} have addressed the issue of  Boltzmann brains \cite{Susskind}, that seem to be in a disagreement with observations. In fact, one would expect to meet a large number of such thermal events along any geodesic, but this does not pose a problem with described measures. Every recyclable bubble contains an infinite number of ordinary observers, while Boltzmann brains are infinitely suppressed once the full probability measure is applied. Similar conclusions are reached in recent publications for the pocket-based measure \cite{Vilenkin2} and volume weighted measure \cite{Linde2}.  

In a recent review paper of different measures of the universe \cite{Aguirre}, an interesting issue of  $L$-tunneling events was discussed. The authors have concluded that the worldline approach would ignore $L$-tunneling events. We disagree with this conclusion. An eternal geodesic that passes through an infinite number of bubbles would undergo $L$-tunneling with probability one, regardless of how unlikely the event is. We conclude that $\bp_s$ and $\bp_e$ prescriptions discussed above do not in principle suffer from the problem of $L$-tunnelings, although we have not done any calculations to include such events. On the other, it is not immediately clear how to include $L$-tunnelings to the recursive measure.

The discussion carried in this article was quite general, however, we have introduced simplified assumptions. The collisions of  bubbles was not considered, which may or may not play a significant role (See Ref. \cite{Guth} for a recent discussion of the issue). The effect of colliding bubbles would not change our prescriptions, but could in principle modify the calculations of all measures. We have also assumed that the bubbles are generated by an irreducible and aperiodic Markov process in order to prove the uniqueness of eternal geodesics (See Appendix). Both of these assumptions seem to be very generic for models of eternal inflation.
 
\section*{Acknowledgments}

It is a pleasure to thank Alex Vilenkin for many stimulating discussions and for a number of useful suggestions. Thanks also goes to Serge Winitzki and Jaume Garriga for helpful comments. This work was supported in part by project ``Transregio (Dark Universe)''.

\section*{APPENDIX}

In this appendix we will show the existence and uniqueness of the single eternal geodesic measure (S measure), by proving the existence and uniqueness of eternal geodesics passing through an arbitrary point $O$.

The concept  of eternal geodesics is closely related to eternal points discussed in Ref. \cite{Serge}. Eternal points are defined only on a given space-like slice ${\cal C}$ of the space-time. The worldline of every observer intersects ${\cal C}$ at a single point, that may or may not be eternal.

\textbf{Definition 1:} A point $E \in {\cal C}$ is called eternal if its world-line never enters a terminal vacuum.

It was shown in Ref. \cite{Serge} that the measure of the set of eternal points is zero. Nevertheless, eternal points always exist if the conditions for eternal inflation are satisfied. We would like to extend this analysis to eternal geodesics. In contrast to eternal points, eternal geodesics can be defined without any reference to a specific space-like slice. 

\textbf{Definition 2:} A geodesic is called eternal, if it entirely lies in recyclable vacua.

Some eternal geodesics remain in the same vacua for entire evolution, while others make transitions from one recyclable vacua to another, but never to a terminal one. However, under the assumption that bubbles are generated by an irreducible (any states is accessible from any  recyclable state) and aperiodic (the return to any state must not occur in some multiple of $k>1$ steps) Markov process, the uniqueness of eternal geodesics can be proved.

\textbf{Theorem 1} {\it (Uniqueness of eternal geodesics):} The abundances of bubbles crossed by eternal geodesics are given by: (a) stationary distribution, (b) independent of initial conditions, and (c) almost surely independent of the geodesic.

\textbf{Proof:} The existence of a stationary distribution independent of initial conditions holds due to the assumption of irreducibility and aperiodicity of the stochastic process. This implies, that there is a unique stationary distribution independent of initial distribution, which corresponds to an eigenvector of a transition matrix with eigenvalue one. This is a special case of the Perron-Frobenius theorem.

Also, due to irreducibility of our process, a randomly chosen eternal geodesic has a zero probability of not crossing some recyclable vacuum. This implies that the distribution of bubbles on eternal geodesics is given by the same Markov process with the largest possible communication class consisting of all recyclable vacua. Thereby, the distribution of bubbles is almost surely independent of the geodesic. This completes the proof of Theorem 1.

To put a measure on the set of geodesics it is convenient to restrict ourselves to all future time-like geodesics passing through a given space-time point $O$. All of these geodesic swap a solid angle of the future light-cone of $O$. The measure of any subset of geodesics is defined as the solid angle swapped by the subset.

Since the probability of an arbitrary observer to remain in  recyclable vacua is zero, the measure of the subset of all eternal geodesics is also zero. Nevertheless, eternal geodesics always exist.

\textbf{Theorem 2} {\it (Existence of eternal geodesics):} For a given space-time point $O$ in recyclable vacua, there always exists an eternal time-like geodesic passing through $O$.

\begin{figure}
\begin{center}
\leavevmode\epsfxsize=3.0in\epsfbox{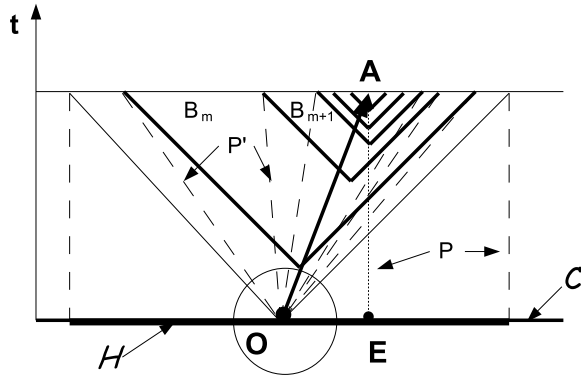}
\end{center}
\caption{Construction of eternal geodesic $OA$ passing through an arbitrary point $O$ in recyclable vacua.}
\label{fig:appendix}
\end{figure}

\textbf{Proof:} Let ${\cal C}$ be an arbitrary space-like surface passing through $O$. Every point $X$ inside of the future light cone of $O$ can be connected by a time-like geodesic passing through $O$. All of these points can also be projected back to $P(X) \in {\cal C}$ by means of the geodesics passing through $X$ and normal to the surface ${\cal C}$. A region ${\cal H} \subset {\cal C}$ swapped by projected geodesic is similar in size to the horizon volume (See Fig.\ \ref{fig:appendix}
). This implies that eternal points almost surely exist within ${\cal H}$ (See Ref. \cite{Serge}). 

Let $E$ be an arbitrary eternal point  in ${\cal H}$. The worldline of $E$ is an eternal geodesic and by Theorem 1 almost surely crosses an infinite number of bubbles of all possible recyclable vacua. The sequence of bubbles $B_n$ is numbered  in the order they are visited by an observer on the geodesic. The worldline of $E$ eventually has to enter the inside region of the future light-cone of $O$, and stay there throughout the evolution. Thus, starting from some $m$ all bubbles lie entirely inside of the light-cone of $O$. We can now define a projection $P'$ of the bubbles by means of geodesics passing through $O$. All of $B_n$ bubbles for $n>m$ lie inside of the future light-cone of $O$, and every next bubble lies inside of all previous bubbles. Therefore, the sequence $P'(B_n)$ forms a projection map of $B_n$'s to the unit three sphere, corresponding to the directions of geodesics originating at $O$, such that $P'(B_{n+1})\subset P'(B_n)$ for all $n>m$. This implies that there exists at least one point $A$ on the unit sphere which lies inside of all projections of bubbles. A geodesic originated at $O$ in the direction corresponding to $A$ by construction intersects an infinite number of bubbles, and therefore is eternal.

\end{document}